\documentclass[useAMS,usenatbib]{mn2e}
\usepackage{times}
\usepackage{graphicx}

\def\aap{A\&A}%
\def\aaps{A\&AS}%
\def\aj{AJ}%
\def\apj{ApJ}%
\def\apjl{ApJ}%
\def\apjs{ApJS}%

\def\iaucirc{IAU Circ.}%
\def\mnras{MNRAS}%
\def\nar{New Astron.~Rev.}%
\def\nat{Nature}%
\def\pasp{PASP}%
\def\qjras{Q.~J.~R. Astron. Soc.}%
\def\ssr{Space Sci.~Rev.}%

\title[Life after eruption. III.]{Life after eruption -- III. 
Orbital periods of the old novae V365 Car, AR Cir, V972 Oph, HS Pup, V909 Sgr,
V373 Sct and CN Vel}
\author[C. Tappert et al.]%
{
C. Tappert,$^{1}$\thanks{E-mail: claus.tappert@uv.cl}
L. Schmidtobreick,$^{2}$
N. Vogt,$^{1}$ and
A. Ederoclite$^{3}$
\footnotemark[1]\thanks{Based on observations with ESO telescopes,
proposal numbers 083.D-0158(A), 086.D-0428(A), 087.D-0323(A), 088.D-0588(A),
089.D-0505(C)}\\
$^{1}$Departamento de F\'{\i}sica y Astronom\'{\i}a, Universidad de 
Valpara\'{\i}so, Avda. Gran Breta\~na 1111, 2360102 Valpara\'{\i}so, Chile\\
$^{2}$European Southern Observatory, Alonso de Cordova 3107, 7630355 Santiago, 
Chile\\
$^{3}$Centro de Estudios de F\'{\i}sica del Cosmos de Arag\'on, Plaza San 
Juan 1, Planta 2, Teruel, E44001, Spain\\
}
\begin{document}

\date{Accepted. Received}

\pagerange{\pageref{firstpage}--\pageref{lastpage}} \pubyear{2013}

\maketitle

\label{firstpage}

\begin{abstract}
We present time-series photometric and spectroscopic data for seven old
novae. They are used to derive the orbital period for the systems
V365 Car (5.35 h), AR Cir (5.14 h), V972 Oph (6.75 h), HS Pup (6.41 h),
V373 Sct (3.69 h), V909 Sgr (3.43 h) and CN Vel (5.29 h). Their addition
increases the number of orbital periods for novae by $\sim$10 per cent.
The eclipsing nature of V909 Sgr is confirmed, and in three other cases (V365 
Car, Ar Cir and V373 Sct) we detect significant photometric orbital 
variability with amplitudes $\ge$0.2 mag in $R$. The resulting period 
distribution is briefly discussed. We furthermore provide new measurements for 
the previously ambiguous coordinates for AR Cir and CN Vel and the 
identification of a new probable W UMa variable in the field of V909 Sgr. The 
spectrum of V972 Oph presents an emission feature redward of H$\alpha$ which 
we tentatively identify with the C{\sc ii} $\lambda\lambda$6578/6583 doublet. 
It is shown that this line originates in the binary and not in a shell, and 
to our knowledge this is the first time that it has been detected in such 
quality in a cataclysmic variable (CV). We argue that this line could be more 
common in CVs, but that it can be easily masked by the broad H$\alpha$ 
emission that is typical for these systems. A closer inspection of the line 
profiles of the other novae indeed reveals an extended red wing in V365 Car, 
CN Vel and AR Cir. In the latter system additionally an absorption counterpart 
blueward of H$\alpha$ is detected and thus in this case a bipolar outflow 
appears as a more likely scenario rather than C{\sc ii} emission.
\end{abstract}


\section{Introduction}
\defcitealias{tappertetal12-1}{Paper I}%
\defcitealias{tappertetal13-1}{Paper II}%

The intrinsic distribution of cataclysmic variables (CVs) is determined by two 
elements. First, there is the period distribution of newly formed CVs, 
i.e.~the distribution of the periods at which the late-type secondary star 
starts to transfer mass on to the white dwarf primary via Roche lobe overflow 
for the first time. Secondly, there is the secular evolution of CVs from this 
period that is governed by continuous angular momentum loss due to magnetic 
braking and gravitational radiation, which decreases the binary separation and 
thus the orbital period for non-degenerate secondary stars 
\citep[e.g.,][]{king88-1}. As CVs that are born with periods $P_\mathrm{orb}
> 3~\mathrm{h}$ evolve below this point the secondary star becomes fully 
convective. It is generally thought that as a consequence magnetic braking 
stops to operate or at least becomes much less efficient, causing the system 
to become detached until it restarts mass transfer at 
$P_\mathrm{orb} \sim 2~\mathrm{h}$ \citep*{rappaportetal83-1}. The result is a 
paucity of CVs with periods between $\sim$2 and 3 h, the so-called period gap. 
Finally, the minimum period at $\sim$80 min represents the turn-around point 
when the secondary star becomes degenerate and the CV now evolves, very 
slowly, back to longer orbital periods due to the inverted mass-radius 
relation. Modelling predicts that most CVs have orbital periods between 
$\sim$80 min and 2 h \citep*{dekool92-1,stehleetal97-1}. 

Although modern surveys have unveiled more of the intrinsic CV 
population \citep[e.g.,][]{gaensickeetal09-2} the currently observed period 
distribution is still affected by a strong observational bias. The latter 
includes a mixture of factors, but the main contributors will be the 
brightness of the CV and frequency and amplitude of the variability. With 
increasing mass-transfer rate $\dot{M}$ the amplitude of the dwarf nova 
outbursts decreases, but the frequency increases, and also the brightness 
of the accretion disc, which is the dominant light source in most CVs 
\citep[e.g.,][and references therein]{lasota01-3}. Since $\dot{M}$ 
for systems above the gap exceeds the one for CVs below it by one to two 
orders of magnitude \citep{townsley+gaensicke09-1}, these are the objects that 
are favoured by the observational bias. 

Classical novae are CVs that undergo a thermonuclear explosion
on the surface of the white dwarf due to the accreted mass having reached
a critical value \citep*{starrfieldetal76-1}. The underlying binary is not
destroyed by the nova eruption and mass transfer is reestablished on 
time-scales not longer than one or two years \citep*[e.g.,][]{retteretal98-2}. 
Nova eruptions can thus be assumed to be recurrent events. The recurrence time 
in classical novae is estimated to be $t_\mathrm{rec} > 10^3~\mathrm{yr}$ 
\citep{sharaetal12-3}. While it is not yet entirely clear how strongly other 
parameters like magnetic fields or the chemical compositions affect 
$t_\mathrm{rec}$, the most important factors have been shown to be $\dot{M}$, 
the internal temperature of the white dwarf $T_c$ and its mass 
$M_\mathrm{WD}$. However, $M_\mathrm{WD}$ is mainly independent of 
$P_\mathrm{orb}$ \citep*{zorotovicetal11-1}, and while 
\citet*{nelsonlaetal04-1} have shown that $T_c$ is an important factor in the 
shaping of the orbital period distribution, \citet{townsley+bildsten04-1} find 
that $T_c$ depends on the long-term average $\dot{M}$. It is thus reasonable 
to assume that the latter represents the overall most important factor for the 
nova period distribution. Since the highest values for $\dot{M}$ are found in 
CVs with $P_\mathrm{orb} \sim 3-4~\mathrm{h}$ \citep{townsley+gaensicke09-1}, 
one would expect that the observed period distribution for novae shows a 
strong maximum at such periods. Detailed calculations by 
\citet{townsley+bildsten05-1} show that roughly 50 per cent of the novae 
should have orbital periods in this range. \citet{diaz+bruch97-1} analyse the 
influence of several observational selection effects, but find that they do 
not strongly affect the overall shape of the period distribution.

The main problem with comparing theoretically derived period distributions
to the actually observed one is that the latter is still significantly
undersampled. The current edition of the \citet{ritter+kolb03-1} catalogue
(update 7.20) lists only 69 post-novae with unambiguous periods. The problem
of undersampling extends to all aspects of post-nova research, which motivated
us to start a project to select candidates for `lost' post-novae via
colour-colour diagrams, to confirm the post-nova spectroscopically, and to
derive the orbital period for suitable, i.e.~sufficiently bright systems
\citep[][hereafter Paper I, II]{tappertetal12-1,tappertetal13-1}. In the
present paper we report on the orbital periods for seven post-novae.

\section{Data handling}
\subsection{Observations and reduction}

\begin{table}
\caption[]{Log of observations.}
\label{obslog_tab}
\setlength{\tabcolsep}{0.15cm}
\begin{tabular}{lllllll}
\hline\noalign{\smallskip}
Object & Date & Grism & $n$ & $t_\mathrm{exp}$ & $\Delta t$ & $R$ \\
       &      & ({\AA})      &     & (s)              & (h) & (mag) \\
\hline\noalign{\smallskip}
V365 Car & 2011 February 26 & \#20  & 3  & 720 & 0.42 & 17.64 \\
         & 2011 February 27 &       & 3  &     & 0.42 & 17.60 \\
         & 2011 February 28 &       & 4  & 900 & 1.59 & 17.57 \\
         & 2011 March 01 &       & 1  &     & --   & 17.61 \\
         & 2011 March 02 &       & 3  &     & 5.69 & 17.67 \\
         & 2011 March 03 &       & 4  &     & 5.83 & 17.64 \\
         & 2012 March 25 &       & 3  &     & 5.17 & 17.61 \\
         & 2012 March 26 &       & 4  &     & 5.11 & 17.73 \\
         & 2012 March 27 &       & 3  &     & 2.26 & 17.67 \\
         & 2012 March 28 &       & 19 &     & 5.45 & 17.68 \\
         & 2012 April 02 &       & 8  &     & 2.46 & 17.61 \\
         & 2012 May 15 &       & 14 &     & 4.06 & 17.69 \\
         & 2012 May 16 &       & 12 &     & 3.54 & 17.66 \\
         & 2012 May 17 &       & 5  &     & 2.01 & 17.69 \\
         & 2012 May 18 &       & 9  &     & 4.80 & 17.70 \\
         & 2012 May 19 &       & 6  &     & 4.06 & 17.68 \\
AR Cir   & 2012 May 17 & \#20  & 6  & 900 & 3.11 & 17.66 \\
         & 2012 May 18 &       & 7  &     & 7.45 & 17.76 \\
         & 2012 May 19 &       & 4  &     & 2.44 & 17.75 \\
V972 Oph & 2011 June 30 & \#20  & 3  & 900 & 3.51 & 15.90 \\
         & 2011 July 01 &       & 7  &     & 9.44 & 15.93 \\
         & 2012 May 16 &       & 2  &     & 3.24 & 15.86 \\
         & 2012 May 17 &       & 2  &     & 0.77 & 15.72 \\
         & 2012 May 18 &       & 5  &     & 6.69 & 15.82 \\
         & 2012 May 19 &       & 5  &     & 2.74 & 15.77 \\
HS Pup   & 2009 May 19 & \#4   & 6  & 3570 & --  & 17.59 \\
         & 2009 May 22 &       & 3  & 1260 & --  & 17.62 \\
         & 2009 May 23 &       & 3  & 1890 & --  & 17.66 \\
         & 2011 February 26 & \#20  & 6  & 720 & 2.07 & 17.89 \\
         & 2011 February 28 &       & 1  &     & --   & 17.96 \\
         & 2011 March 01 &       & 3  & 900 & 2.21 & 17.88 \\
         & 2011 March 02 &       & 3  &     & 1.83 & 17.92 \\
         & 2011 March 03 &       & 5  &     & 4.15 & 18.01 \\
V909 Sgr & 2012 May 16 & \#20  & 9  & 900 & 5.50 & 20.15 \\
         & 2012 May 17 &       & 6  &     & 1.62 & 20.03 \\
         & 2012 May 18 &       & 7  &     & 5.96 & 20.11 \\
         & 2012 May 19 & $V$   & 175 & 60 & 4.74 & 20.46$^a$ \\
V373 Sct & 2011 June 29 & \#20  & 4  & 900 & 6.40 & 18.82 \\
         & 2011 June 30 &       & 4  & 900 & 5.82 & 18.83 \\
         & 2011 July 01 &       & 5  & 900 & 6.94 & 18.63 \\
CN Vel   & 2009 May 19 & \#4   & 6  & 2160 & --  & 17.48 \\
         & 2009 May 22 &       & 3  & 540  & --  & 17.49 \\
         & 2009 May 23 &       & 3  & 540  & --  & 17.43 \\
         & 2009 May 24 &       & 3  & 2100 & --  & 17.45 \\
         & 2011 February 26 & \#20  & 3  & 750 & 0.44 & 17.56 \\
         & 2011 February 28 &       & 1  & 900 & --   & 17.58 \\
         & 2011 March 01 &       & 3  & 900 & 3.28 & 17.53 \\
         & 2011 March 02 &       & 2  & 900 & 1.59 & 17.53 \\
         & 2011 March 03 &       & 4  & 900 & 5.89 & 17.54 \\
\hline
\multicolumn{7}{l}{a) $V$ magnitude}\\
\end{tabular}
\end{table}

The time-series data were obtained on several observing runs in 2011 and 
2012 at the ESO-NTT, La Silla, Chile, using EFOSC2 \citep*{eckertetal89-1}. 
The meteorological conditions were not photometric, with the 2011 runs being
strongly affected by clouds and poor seeing, while for the 2012 runs the
sky was mostly clear with reasonably good seeing $\sim1.0$ arcsec.
Spectroscopy was performed with grism \#20 and a 1 arcsec slit, 
yielding a wavelength range 6035--7135 {\AA} at a spectral resolution of 3.8 
{\AA}, the latter measured as the full width at half-maximum (FWHM) of an arc 
line. Additional low-resolution spectroscopy with grism \#4 and a 
1 arcsec slit was taken in 2009 May for some of the systems. The corresponding 
wavelength range was 4000--7400 {\AA} at a resolution of 11 {\AA}. The 
acquisition of the targets was done by taking 10 or 20 s exposures of the 
field with a Bessel $R$ filter (ESO \#642). For one object (V909 Sgr), 
additionally a light curve was taken using a Bessel $V$ filter (ESO \#641). A 
detailed observing log is given in Table \ref{obslog_tab}, stating the 
observing dates, the grism used, the number $n$ of individual exposures in a 
specific night, the exposure time $t_\mathrm{exp}$ (for the grism \#4 data 
this is the total exposure time), the time range $\Delta t$ covered by the 
time-series observations and the average $R$-band magnitude during that
night. For the photometric time-series data of V909 Sgr, the average $V$ 
magnitude is given.

Data reduction was performed with {\sc iraf}. For the photometric data
(the acquisition frames and the light curve for V909 Sgr), this included
only the subtraction of bias frames, but no flat-field correction because 
EFOSC2 flats are affected by a central light concentration. Photometric
magnitudes for all stars on the frames were extracted using {\sc iraf}'s 
{\sc daophot} package and the standalone {\sc daomatch} and {\sc daomaster} 
routines \citep{stetson92-1}. The aperture radius for a given frame was chosen 
as a few tenths of a pixel less than the average FWHM of the stellar 
point-spread function on that frame. Differential magnitudes were computed
with respect to the average of suitable comparison stars in the neighbourhood 
($\pm$300 pixels) of the target's position and within $-$0.5 to $+$1.5 mag of 
its brightness. Calibrated magnitudes were determined by
comparison to previously taken data or to the $R_F$ magnitudes from the
{\em Hubble Space Telescope} Guide Star Catalogue 
\citep[GSC, ][]{laskeretal08-1} as provided by the ESO
archive. The $R_F$ passband differs significantly from the Bessel $R$
filter curve, leading to systematic offsets between the two magnitudes
depending on the spectral energy distribution of the object. From our
comparison to several field stars, we estimate a typical uncertainty 
of $\pm0.2$ mag regarding this offset.
We derive corresponding values for the $V$ values for the comparison with
earlier observations by correcting for the $V\!-\!R$ colour derived
elsewhere. This introduces an additional uncertainty, because changes in
brightness could well be accompanied by changes in colour (e.g., a brighter
accretion disc is usually also bluer). However, with respect to the brightness
differences observed in the present work and the previously reported colours of
novae, e.g.~by \cite{szkody94-2} and \citep{zwitter+munari95-1,%
zwitter+munari96-1}, we do not expect such change
to exceed $\sim$0.1 mag. 

Reduction of the spectroscopic data consisted of the subtraction of the
bias frame and the division by a flat-field that was normalized by fitting a
high-order cubic spline to it. The spectra were extracted using the optimal
extraction algorithm \citep{horne86-1} as implemented in {\sc iraf}'s
{\sc onedspec} package. Wavelength calibration was achieved with the
data from HeAr lamps. The data were not flux calibrated, but a correction of 
the instrumental response function that was established from previous 
observations was applied to the grism \#4 spectra, so that the continuum trend 
should be mostly reproduced correctly (the data were not taken at airmasses 
higher than 1.3).

\subsection{Analysis}

A large number of our objects show comparatively weak 
and broad emission lines. Additionally, due to the faintness of the objects 
and the necessarily short exposure times, the spectroscopic data have a low 
signal-to-noise ratio (S/N) typically not higher than $\sim10$. This presents 
significant difficulties for measuring the radial velocity displacement of the 
H$\alpha$ emission line by fitting a Gaussian function to it. We 
therefore adopt a different strategy by computing the average spectrum and
iteratively shifting and scaling that spectrum until it yielded a visually
reasonable fit to the individual spectra. This `eyeball cross-correlation'
has the disadvantage of being considerably more time-consuming, but it yields 
in general smoother radial velocity curves because the human eye can more 
easily ignore the influence of noise or additional emission components 
\citep[e.g., ][]{tappertetal03-1} than a mathematical function. A few spectra 
(e.g.~for AR Cir) that present more Gaussian line profiles and good S/N
allowed for a comparison of the two methods, and we found the respective 
results to agree well with each other. Finally, to account for possible 
instrumental flexure, such determined radial velocities are corrected for the 
actual position of the [O{\sc i}] $\lambda$6363.78 {\AA} night sky line.

The radial-velocity and light curves are examined for periodicities using
the \citet{scargle82-1} and analysis-of-variance 
\citep[AOV;][]{schwarzenberg-czerny89-1} algorithms as implemented in
{\sc eso-midas} \citep{warmels92-1}. From our experience we find that the
former is useful to restrict the range of possible periods, while the latter
more successfully suppresses alias periods within this range. The uncertainty
of the selected period was estimated from the width of the peak in the Scargle
periodogram, since their shape very roughly resembles a Gaussian, while the
AOV peaks are more irregular.

The radial velocities are subsequently folded with the selected period and 
fitted with a sinusoidal function
\begin{equation}
v_r (\varphi) = \gamma - K_\mathrm{H\alpha} \sin (\varphi-\varphi_0),
\label{rv_eq}
\end{equation}
where $\gamma$ is the constant term, $K_\mathrm{H_\alpha}$ the semi-amplitude 
and $\varphi$ is the orbital phase with respect to $\varphi_0$. The time of 
this zero-point of the orbital phase is determined either as the red-to-blue 
crossing of the radial velocity curve, and thus corresponds to the superior 
conjunction of the emission source, or with respect to orbital features (e.g., 
an eclipse) detected in the photometric light curve. The errors associated 
with these parameters have been estimated by the means of a Monte Carlo 
routine.

\section{Results}

\begin{figure*}
\includegraphics[width=1.8\columnwidth]{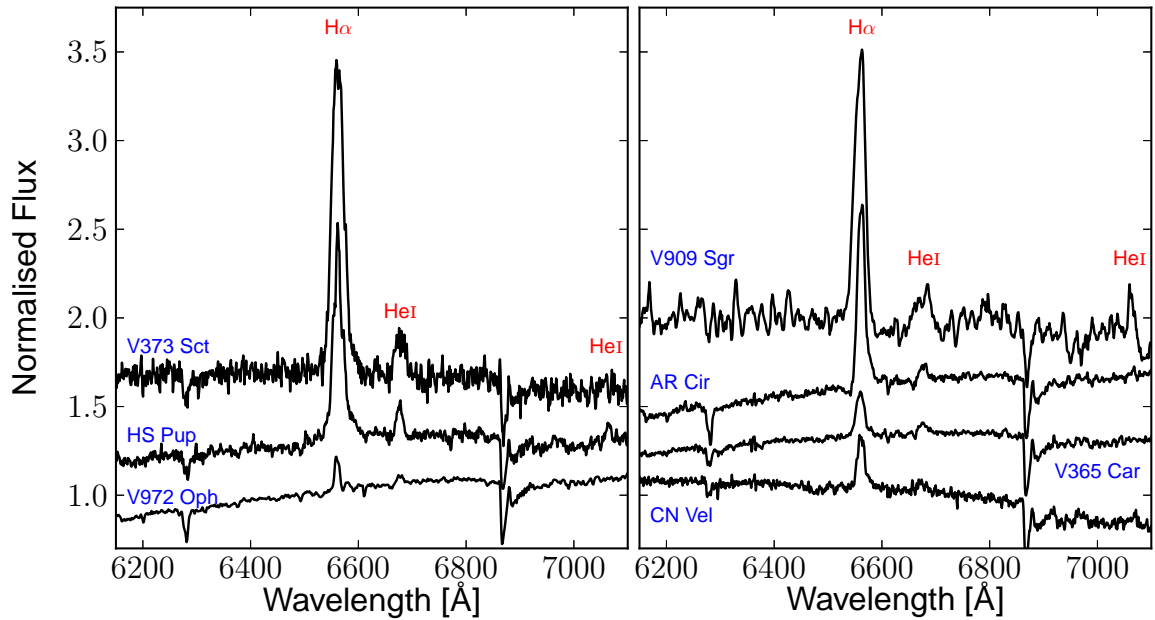}
\caption[]{Average spectra for the seven novae as labelled. The spectra have 
been normalized by dividing through the mean flux and vertically displaced for 
display purposes. The V909 Sgr data have been smoothed with a 3$\times$3 box 
filter.}
\label{allspec_fig}
\end{figure*}

\begin{table}
\caption[]{Radial velocity parameters as defined in equation (\ref{rv_eq}).}
\label{rvpar_tab}
\begin{tabular}{llll}
\hline\noalign{\smallskip}
Object & $\gamma$ (km s$^{-1}$) & $K_\mathrm{H\alpha}$ (km s$^{-1}$) 
& $\sigma_{\varphi_0}$ (orbits)\\
\hline\noalign{\smallskip}
V365 Car 2011 & $-$16(10)  & 103(15) & 0.021 \\
V365 Car 2012 & $-$25(5)   & 116(7)  & 0.010 \\
AR Cir        & 21(3)      & 101(4)  & 0.006 \\
V972 Oph 2011 & $-$3(4)    & 115(7)  & 0.007 \\
V972 Oph 2012 & $-$79(4)   & 104(7)  & 0.007 \\
HS Pup        & 85(4)      & 119(5)  & 0.006 \\
V909 Sgr      & $-$117(11) & 109(15) & 0.021 \\
V373 Sct      & 55(5)      & 188(8)  & 0.005 \\
CN Vel        & 7(5)       & 156(6)  & 0.008 \\
\hline\noalign{\smallskip}
\end{tabular}
\end{table}

\subsection{V365 Carinae = Nova Car 1948}

\begin{figure}
\includegraphics[width=\columnwidth]{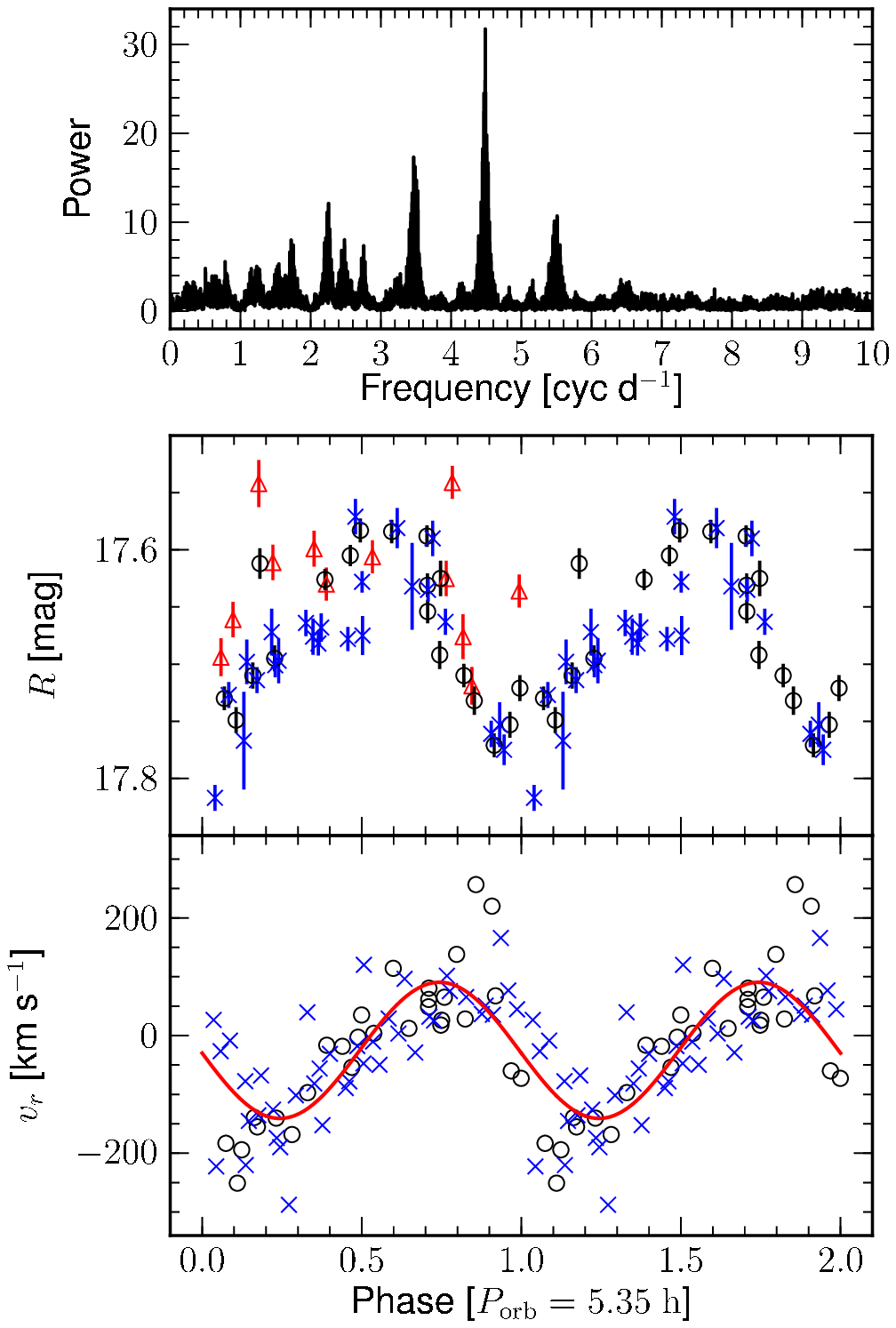}
\caption[]{Time-series data on V365 Car. Top: AOV periodogram of the 2012
photometric data. Middle: light curves from 2011 February (triangles),
2012 March (circles) and May ($\times$). Bottom: radial velocities from
2012 March (circles) and May ($\times$).}
\label{v365cardata_fig}
\end{figure}

This star was never observed `real time' during its nova eruption. Its remnant
was detected about 26 years after its outburst maximum during a search for 
emission line stars, assigning it the identification He 3-558 due to its 
H$\alpha$ emission.  \citet{henize+liller75-1} noted that the spectrum of this 
object resembles that of an old nova, and carried out a search for 
variability in the Harvard plate archive. They detected that, in fact, it had 
erupted as a nova with a maximum photographic brightness 
$m_\mathrm{pg} = 10.1$ mag in 1948 September 25  and a very slow decline 
afterwards \citep[$t_3 = 530~\mathrm{d}$ according to][]{duerbeck87-1}. 
\citet{gill+obrien98-1} performed a search for a nova shell without success. 
In a spectrum taken in 1995, \citet{zwitter+munari96-1} find the 
Bowen/He{\sc ii} emission blend to be significantly stronger than the Balmer 
emission lines, which they interpret
as the signature of a rather hot white dwarf. The continuum, on the other hand,
presents only a modest blue slope, but the authors speculate on this being
due to interstellar reddening. They furthermore note the absence of He{\sc i} 
emission. The brightness at that time was $V$ = 18.31 mag, thus yielding a 
comparatively small eruption amplitude $\Delta m \sim 8~\mathrm{mag}$ as
expected from a slow nova \citep[e.g.,][]{duerbeck81-1}. V365 Car
was observed at the same brightness level by \citet{woudt+warner02-3}, who 
performed high speed photometry during three nights in 2000 March for a total 
of 21 h. Their most extended runs present a hump-like wave structure with an 
amplitude of about  0.15 mag. They concluded that the corresponding period 
should be either 6.86 or 8.00 h. A slow flickering with $\sim$0.1 mag 
amplitude was also present. V365 Car is located near the outer border of the 
open cluster NGC 3532, but located beyond it, because the distance of this 
nova is of the order of 3.5 kpc, a factor 10 larger than that of the cluster 
\citep{henize+liller75-1}. 

We observed V365 Car on three occasions, in 2011 February, 2012 March and
May. Comparing the acquisition frames with the GSC we find as average
values $R$ = 17.63(05), 17.67(06) and 17.68(06) mag for the runs in
above sequence. The values in the parentheses here do not indicate a
measurement uncertainty, but correspond to the mean variability. The 2012
runs thus show the object at the same brightness level, while it appears 
slightly brighter in the 2011 data. \citet{zwitter+munari96-1} give $V-R_{C} = 
0.43~\mathrm{mag}$ for a Cousins $R$ filter curve, yielding 
$V \sim 18.1~\mathrm{mag}$ for our data. Since the involved uncertainties
amount to $\sim$0.2 mag this does not necessarily mean that the object was
in a (much) different brightness state than during the 
\citet{zwitter+munari96-1} and the \citet{woudt+warner02-3} observations. 

The average spectrum, composed out of the data of all runs, is given in
Fig.~\ref{allspec_fig}. 
The main feature is a comparatively weak H$\alpha$ line with an equivalent
width $W_\mathrm{H\alpha} = 5~\mathrm{\AA}$\footnote{For convenience, we use 
here and throughout the paper positive values for $W_\lambda$. This is 
unambiguous since we exclusively refer to emission lines.}.
We also note the presence of the He{\sc I} $\lambda$6678 emission line that 
was not detected in the \citet{zwitter+munari96-1} spectrum. Looking at their 
data suggests that it was probably hidden by the low S/N and the lower 
spectral resolution. 

The combined 2012 time-series photometric data from the acquisition frames
yield a strong and unambiguous periodic signal at a frequency
$f = 4.45~\mathrm{cycle~d^{-1}}$ (top plot in Fig.~\ref{v365cardata_fig}).
This peak is equally present in the combined 2012 radial velocities (although
accompanied by an important alias at $5.45~\mathrm{cycle~d^{-1}}$), while the
2011 data do not give any useful result due to the low S/N and insufficient
sampling. Since folding the data according to above frequency yields reasonable
curves for all data sets, we identify this signal as corresponding to an
orbital modulation at $P_\mathrm{orb} = 0.2247(40)~\mathrm{d} 
= 5.35(10)~\mathrm{h}$. 

In the middle plot of Fig.~\ref{v365cardata_fig}, we include the 2011 
light-curve data (for clarity only one orbit is plotted) to show that these in 
general present the same modulation as the 2012 data, but are systematically
offset to brighter magnitudes. The shape of the $R$ light curve is that of a
sinusoid or a hump with a semi-amplitude of $\Delta m_R \sim 0.1~\mathrm{mag}$.
The origin of the variation could therefore be either irradiation of the 
secondary star by the primary or the bright spot that is formed by the 
accretion stream from the secondary star hitting the accretion disc around the 
primary. The low amplitude, together with the stochastic variations caused
by flickering, makes it impossible to distinguish between these two scenarios 
on the basis of our data, although the presence of flickering rather favours
a bright-spot possibility. The zero-point of the radial-velocity curve defined 
as the superior conjunction of the emission source coincides with the minimum 
of the light curve. We use this zero-point to define the ephemeris of the 
variation as
\begin{equation}
T_0 ({\rm HJD}) = 245\,6067.4692(23) + 0.2247(40)~E,
\end{equation}
where $E$ is the cycle number. The radial-velocity parameters are summarized
in Table \ref{rvpar_tab}, the last column there giving the error associated
with the determination of the zero-point $\varphi_0$.

\subsection{AR Circinis = Nova Cir 1906}
\label{arcir_sec}

\begin{figure}
\includegraphics[width=\columnwidth]{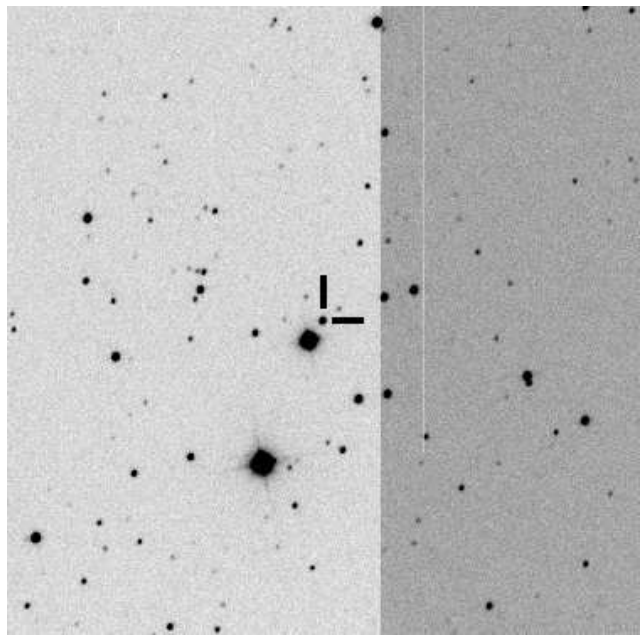}
\caption[]{$R$-band finding chart for AR Cir. The size is 1.5$\times$1.5
arcmin$^2$. North is up, east is to the left.}
\label{arcirfc_fig}
\end{figure}

\begin{figure}
\includegraphics[width=\columnwidth]{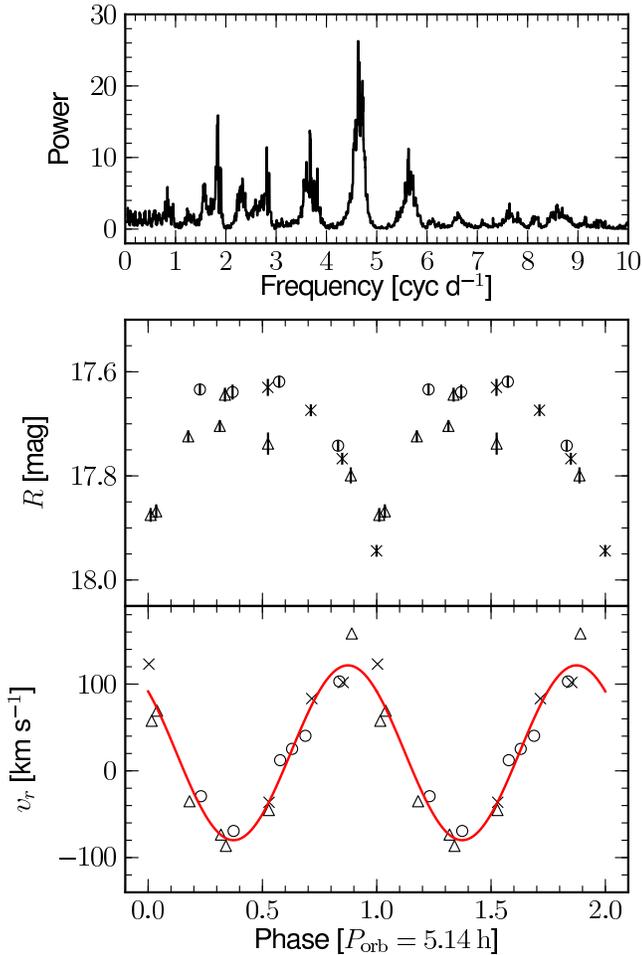}
\caption[]{Time-series data on AR Cir. Top: AOV periodogram of the radial
velocities. Middle: $R$ light curve. Bottom: radial velocities. In the
latter two, different symbols represent different nights.}
\label{arcirdata_fig}
\end{figure}

\begin{figure}
\includegraphics[width=0.7\columnwidth]{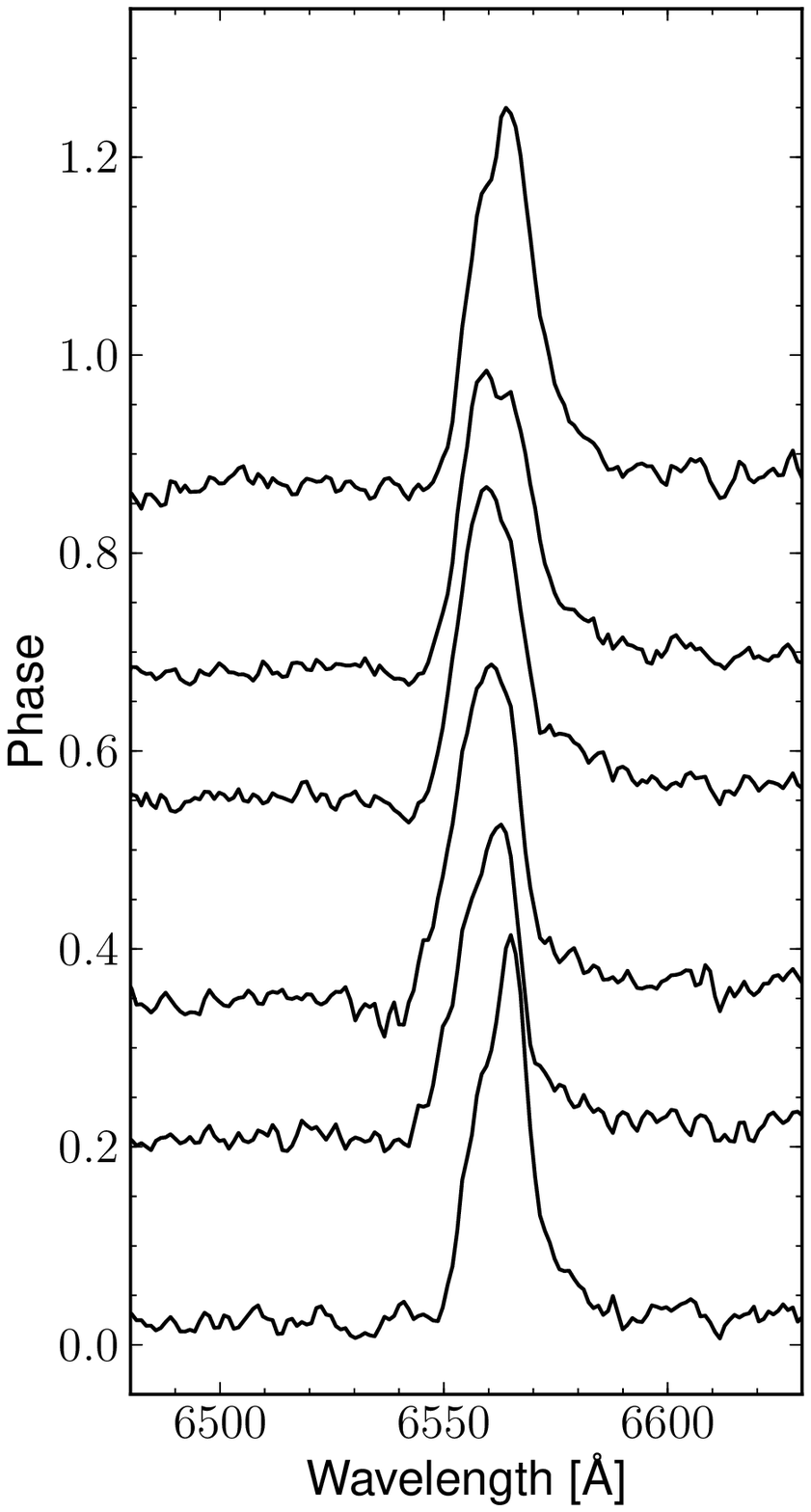}
\caption[]{Binned spectra of AR Cir vertically offset according to orbital 
phase.}
\label{arcirlprof_fig}
\end{figure}

\citet{pickering07-54} announced the detection of this nova on Harvard patrol 
plates and gave a photographic maximum brightness $m_\mathrm{pg} =  
9.5~\mathrm{mag}$. \citet{duerbeck87-1} cites a number of references for 
the long-term light curve and derives $t_3 = 415~\mathrm{d}$. 
\citet{duerbeck+grebel93-1} reported the system to have a close visual K3V 
companion of $V$ = 14 mag, with a separation of only 3.1 arcsec from 
AR Cir. A corresponding correction for the contribution of the companion of 
the nova eruption light curve yielded a maximum photographic magnitude
10.5, and the quiescent magnitudes and colours $V$ = 18.31, $B-V$ = 1.25 and 
$V-R$ = 1.0. Like V365 Car, AR Cir is thus a member of the novae that have a
small eruption amplitude and show a slow decline. The spectrum taken by
\citet{duerbeck+grebel93-1} shows a 
prominent H$\alpha$ emission line on a reddened continuum. 
\citet{gill+obrien98-1} performed a search for a nova shell without success. 
The available infrared photometric data on AR Cir are erroneous, because the
values obtained by \citet{harrison92-1} refer to the combined light of
the nova and its close visual neighbour, while \citet{saitoetal13-2} measure
only the latter (this will be corrected in the upcoming electronic version
of the catalogue; Saito, private communication).

Because of the above mentioned source confusion, and because the 
\citet{downesetal05-1} catalogue lists the coordinates of the visual companion 
instead of the nova, we have measured accurate (to $\sim$0.16 arcsec) 
coordinates from our acquisition frames using Starlink's 
{\sc gaia}\footnote{\tt http://astro.dur.ac.uk/$\sim$pdraper/gaia/gaia.html} 
tool (version 4.4.1) in combination with the UCAC3 \citep{zachariasetal10-1} 
catalogue \citepalias[see also][]{tappertetal12-1}. Such measured position
is
\begin{equation}
\alpha_\mathrm{2000.0} = 14\!:\!48\!:\!09.26,~~
\delta_\mathrm{2000.0} = -60\!:\!00\!:\!24.8.
\end{equation}
Fig.~\ref{arcirfc_fig} presents the corresponding finding chart.

From our photometry we find a mean magnitude $R$ = 17.73(10) mag, with the
value in the parentheses giving the mean variation. Taking $V-R$ = 1.0 from
\citet{duerbeck+grebel93-1} this yields $V \sim 18.7$. Taking into account
the estimated uncertainty in our comparison to GSC magnitudes of $\sim$0.2 mag,
the object still can be suspected to be a few tenths of a magnitude fainter 
during our observations.

The average spectrum is given in Fig.~\ref{allspec_fig}. 
The H$\alpha$ emission line ($W_\mathrm{H\alpha} = 19~\mathrm{\AA}$) and the 
He{\sc i} line at 6678 {\AA} (and to a lesser extent also 7065 {\AA}) can be
clearly identified. In spite of the red continuum slope, the spectrum
does not show any obvious absorption features from the secondary star.
Consulting NASA's Infrared Science Archive (IRSA) web interface that provides 
the interstellar extinction as measured by \citet*{schlegeletal98-2}, we find 
$E(B\!-\!V) = 5.3 \pm 0.3$ for AR Cir. This represents by far the largest 
reddening within our sample, and it is thus reasonable to assume that the 
observed red continuum is mainly due to reddening and does not represent the 
manifestation of the secondary star. Testing different values for the 
reddening we find that a correction for $E(B\!-\!V) \sim 2$ yields a similar 
continuum slope as for the other novae. A more precise measurement would 
require the analysis of interstellar absorption lines.

Applying the AOV algorithm to the radial velocities yields a peak 
corresponding to a period $P = 0.2143(72)~\mathrm{d} = 5.14(17)~\mathrm{h}$
(Fig.~\ref{arcirdata_fig}, top). The peak is comparatively broad due to
the uneven sampling of the data, but folding the radial velocities on this
period yields a smooth sinusoid with a standard deviation of 
$\sigma = 18~\mathrm{km~s^{-1}}$ (Fig.~\ref{arcirdata_fig}, bottom). The
parameters of the best fit to the data are given in Table \ref{rvpar_tab}.
The $R$-band light curve shows a variation with a semi-amplitude of 
$\sim$0.16 mag consisting of a broad maximum and a comparatively sharp
minimum. This resembles a (partial) eclipse of the light source rather than
an orbital hump or a sinusoidal variation. We use this minimum to define
the ephemeris by computing a polynomial function of the second order to the
eclipse shape, yielding
\begin{equation}
T_0 ({\rm HJD}) = 245\,6067.59969(06)+0.2143(72)~E.
\end{equation}

The phase difference between the
photometric minimum and the zero-point of the H$\alpha$ radial-velocity curve 
amounts to 0.12 orbits. Thus, in a coordinate system that has its point of
origin in the centre-of-mass and whose axis connects the eclipsed continuum
source (at $0^\circ$) with the eclipsing source (at $180^\circ$), the H$\alpha$
emission source is located at $\sim$43$^\circ$ (clockwise). Identifying the
eclipsing source with the secondary star implies a rather unusual location
for the H$\alpha$ emitter. A bright spot, for example, would be expected at 
$\sim$135$^\circ$. However, the bright-spot scenario is still possible if
the accretion disc is small and the mass ratio is large so that the secondary
star and the bright spot are on opposite sides of the centre-of-mass. To
confirm this one would need radial velocities from the secondary star. If the
disc indeed is small such should be obtainable with time-series infrared
spectroscopy.

Finally, we find that the H$\alpha$ line profile shows an intriguing 
variability. In Fig.~\ref{arcirlprof_fig} we plot binned spectra (three spectra
per bin with the exception of the bin corresponding to phase 0.21, which 
includes only two). 
We can identify four contributors to the H$\alpha$ line profile: the
main component that makes up the base of the line and probably originates
in the accretion disc, a narrow peak that moves within this base at 
slightly higher velocities and in CVs usually represents emission from
the region of the bright spot or from the secondary star, a weak blue 
absorption component that is most pronounced in the spectra around phase 0.5, 
and finally a weak red emission component that can be detected in all bins, but 
again is best visible at the same time as the absorption component.
This simultaneous presence of a high-velocity blue absorption with a red 
emission counterpart indicates that these components represent redshifted and 
blueshifted H$\alpha$ caused by an optically thick wind or outflowing material 
similar to a P\,Cyg profile. Measuring the separation between the absorption 
and emission P Cyg component with a cursor yields a rough estimate of the 
(projected) outflow velocity to 700 $\pm$ 100 km s$^{-1}$.

\subsection{V972 Ophiuchi = Nova Oph 1957}
\label{v972oph_sec}

\begin{figure}
\includegraphics[width=\columnwidth]{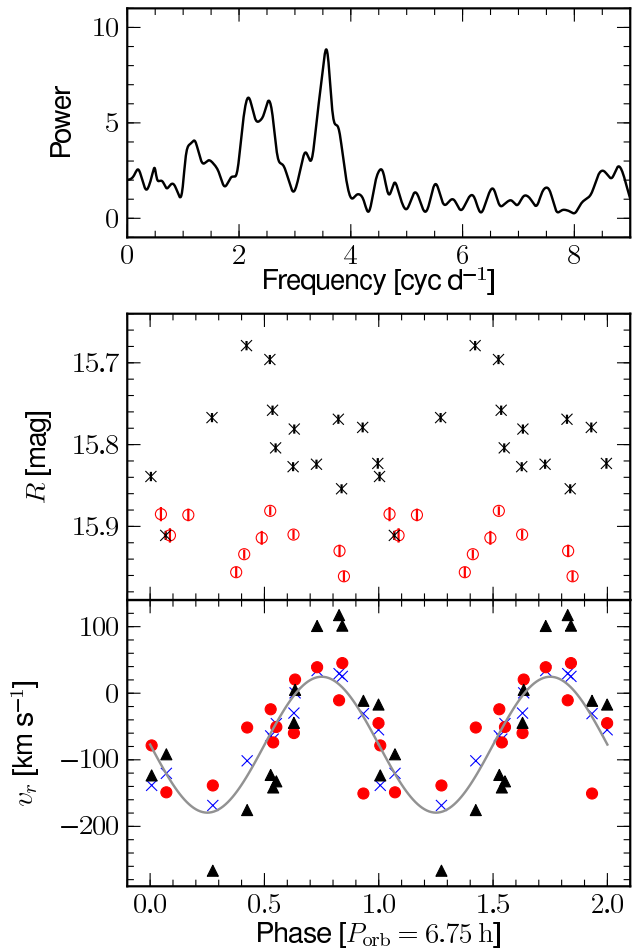}
\caption[]{Time-series data on V972 Oph. Top: combined Scargle periodogram. 
Middle: $R$ light curves from 2011 June/July (circles) and from
2012 May ($\times$). Bottom: radial velocities of the H$\alpha$ line 
($\times$), its additional emission component (triangles), and the line
tentatively identified as C{\sc ii} (circles). The sine represents the best
fit to the H$\alpha$ data.}
\label{v972ophdata_fig}
\end{figure}

\begin{figure}
\includegraphics[width=0.7\columnwidth]{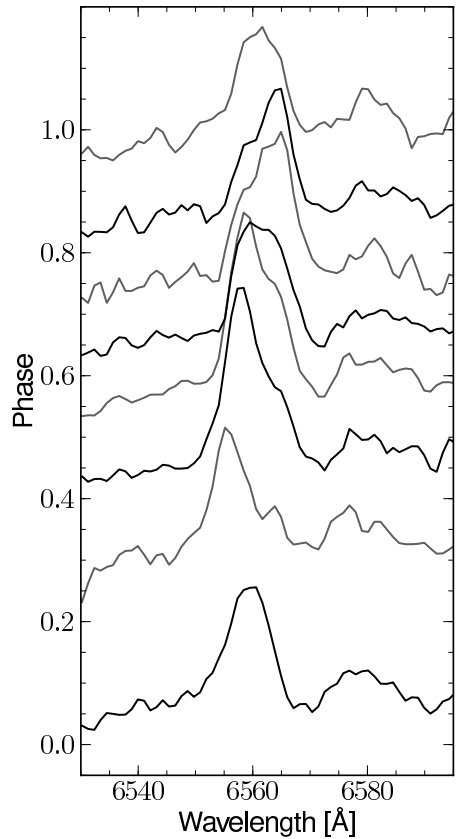}
\caption[]{Close-up of the spectral region around H$\alpha$ plotted versus
orbital phase for V972 Oph. Spectra within the same 0.1 orbits wide phase bins
have been averaged.}
\label{v972ophbins_fig}
\end{figure}

Detected by Haro in 1957 on Tonantzintla plates \citep{haro+whipple58-1}, this 
nova reached $m_\mathrm{pg}$ =  8.0 mag at maximum. \citet{duerbeck87-1} gives
$t_3 = 176~\mathrm{d}$ and reports a pre-maximum halt of at least 50 d.  
\citet*{ringwaldetal96-3} present a continuum spectrum between 4000 and 10000 
{\AA}, with a single weak H$\alpha$ emission line, but with a rather 
strong continuum luminosity excess between 4000 and 5000 {\AA}. This excess is 
not present in \citet{zwitter+munari96-1} who find a comparatively steep red 
continuum, strong He{\sc ii} $\lambda$4686 emission, H$\beta$ as a narrow 
emission line within an absorption trough, and H$\alpha$. The authors remark 
on the absence on He{\sc i} emission, and conclude that the system harbours a 
hot white dwarf, but in fact the He{\sc i} $\lambda$6678 emission line is 
clearly present in their spectrum. Due to the absence of late-type absorption 
features they suggest that the companion has a spectral type F or earlier, 
requiring considerable interstellar extinction to explain the red continuum. 
\citet*{woudtetal04-2} took high-speed photometry on three occasions for a 
total of 8.5 h. They find some flickering activity with a maximum amplitude of 
0.1 mag at a time-scale of about 15 min, but no longer periodic modulation.
The two available infrared studies 
\citep[2MASS and VVV,][respectively]{hoardetal02-1,saitoetal13-2} differ 
by about 4 mag. A comparison of the coordinates shows that the Two Micron All
Sky Survey (2MASS) data are the correct ones, while the object is 
misidentified in the VVV. V972 Oph appears to show some long-term low-amplitude
variations: \citet{ringwaldetal96-3} report $V$ = 16.7 mag in 1991 mid-June; 
\citet{zwitter+munari96-1} find the star at $V$ = 16.56 mag in 1994 April 18;
and \citet{woudtetal04-2} measure $V$ = 15.9 mag in 2002 April 2, $V$ = 16.1
mag in 2003 August 29 and $V$ = 16.2 mag one night later.

We observed the system on two occasions: on two nights in 2011 June/July
and on four consecutive nights in 2012 May (Table \ref{obslog_tab}). From
comparison of the acquisition frames with the GSC we find that the object
on average was slightly brighter during the 2012 run, with $R \sim$15.8 mag,
than on the 2011 run, with $R \sim$15.9 mag. \citet{ringwaldetal96-3}
and \citet{zwitter+munari96-1} measured similar colours, $V\!-\!R$ = 0.7 mag 
and 0.63 mag, respectively. For our data, this yields $V \sim$16.5
and $\sim$16.6 mag. This suggests that V972 Oph was in a brightness state
close to that during the two studies above, and slightly fainter than during
the observations by \citet{woudtetal04-2}.

The time span between the two runs is too large to combine the data for
a time-series analysis, so the sets had to be analysed individually.
The periodograms in each case present two significantly strong peaks, but
only the one at $f = 3.558~\mathrm{cycle~d^{-1}}$ is common to both sets
(Fig.~\ref{v972ophdata_fig}, top). Folding the data with this frequency 
yields a smooth sine curve (Fig.~\ref{v972ophdata_fig}, bottom), and we thus 
identify the modulation with an orbital one corresponding to a 
period $P_\mathrm{orb} = 0.281(8)~\mathrm{d} = 6.75(19)~\mathrm{h}$. Sine fits 
to the radial velocities yield the parameter collected in Table 
\ref{rvpar_tab}. Defining the zero-points of the phase as the time of
the red-to-blue crossing of the radial-velocity curves yields the
ephemeris
\begin{equation}
T_0 ({\rm HJD}) = 245\,5744.719(02) + 0.281(8)~E
\end{equation}
for the 2011 June/July set, and
\begin{equation}
T_0 ({\rm HJD}) = 245\,6067.749(02) + 0.281(8)~E
\end{equation}
for the 2012 May set. As we will discuss below, there are significant
differences in the light curves of the two runs that could
in principle also affect the profile of the emission line. The two ephemerides
thus do not necessarily correspond to the same physical zero-point of the
orbital motion.

The average spectrum shows a peculiar broad line centred at $\sim$6580 {\AA}.
The line moves approximately parallel with the H$\alpha$ line; a sine fit
to its radial velocity yields a semi-amplitude $K_\mathrm{6580} = 
78(11)~\mathrm{km~s^{-1}}$, which is $\sim$2.4 $\sigma$ lower than 
$K_\mathrm{H\alpha}$,
and a phase difference of 0.04(02) orbits for the 2012 May data
(Fig.~\ref{v972ophdata_fig}, bottom). For 
comparison, for the He{\sc i} $\lambda$6678 emission line we find 
$K_{6678} = 119(9)~\mathrm{km~s^{-1}}$ and a phase difference of 0.06(01).
We can discard the possibility that the two lines in reality form one broad 
H$\alpha$ line that is split by an absorption component, because the bluer 
line of the two clearly shows the presence of an additional component that 
moves within that blue line and never crosses over to the red one. In fact, 
some of the phase-binned spectra shown in Fig.~\ref{v972ophbins_fig} hint at a 
corresponding emission component being present in the red line. The blue line
thus necessarily must be H$\alpha$. The identity of the red line is less clear.
Because it shows the same orbital motion as the other lines it can not
be a forbidden line originating in a shell, e.g.~[N{\sc ii}] 
$\lambda$6584\footnote{An additional argument against this possibility is that
the shell was not detected in H$\alpha$/[N{\sc ii}] imaging conducted by
\citet{gill+obrien98-1}.}. Examining the identification list of lines in
stellar spectra \citep{coluzzi99-1} we find as the most likely possibility
that the line represents the C{\sc ii} $\lambda\lambda$6578/6583 doublet.
Carbon emission lines have been observed previously, e.g. in the post-novae
V840 Oph \citep{schmidtobreicketal03-5} and CP Pup \citep{bianchinietal12-1}.
However, both these systems show considerable C and He{\sc ii} emission in the
blue ($<$6000 {\AA}) wavelength range, which does not appear to be the case in 
V972 Oph, although the available blue spectra 
\citep{zwitter+munari96-1,ringwaldetal96-3} have low S/N and/or low spectral
resolution and can thus provide no concluding evidence. Additionally, strong
C{\sc ii} $\lambda\lambda$6578/6583 emission has been reported for the 
helium nova V445 Pup in the first few months after eruption, but there they 
are clearly related to the outflow \citep{wagneretal01-4,iijima+nakanashi08-1}. 
We note that the separation between the suspected C{\sc ii} line and 
H$\alpha$ in velocity units amounts to $\sim$700 km s$^{-1}$, i.e.~the same
as the suspected outflow in AR Cir (Section 
\ref{arcir_sec}). However, in V972 Oph we do not find any evidence for an
accompanying blue absorption or emission component\footnote{As can be seen
in the average spectrum in Fig.~\ref{allspec_fig} there is an absorption
line at $\sim$6530 {\AA}, but apart from having about twice the separation from
H$\alpha$ as the suspected C{\sc ii} emission, it also does not show any 
detectable velocity variations and is thus probably of interstellar origin.}.
Considering the weakness of the H$\alpha$ emission it seems unlikely that
such an absorption component could be hidden in its wings.

Folding the photometric $R$ data on the spectroscopic period shows that,
in spite of the small offset of $\sim$0.1 mag in average magnitude between 
the 2011 data and the 2012 data, the light curves present significantly 
different shapes that are likely to reflect structural difference at the time 
of the two observing runs (Fig.~\ref{v972ophdata_fig}). The 2011 light curve 
is the slightly fainter one and is reminiscent of an ellipsoidal variation with
minima at spectroscopic phases $\sim$0.3 and $\sim$0.9. In contrast, the 2012 
data show a hump at phase $\sim$0.5 and a minimum at $\sim$0.1 orbits. The
latter, however, is defined by only one data point and therefore has to
be treated with some caution. 
Additionally, one has to take into account that both data sets have been 
folded with respect to their individual spectroscopic ephemeris, and as 
remarked above these do not necessarily refer to the same orbital 
configuration. 

With the available data it is not possible to say if the differences in the
photometric behaviour are also reflected in the line emission distribution.
The 2011 spectra have much lower S/N and are more affected by 
fringing, so that an examination of the line profile does not provide
further information. The average spectra, at least, show identical line
strengths (e.g., $W_\mathrm{H\alpha} \sim 2~\mathrm{\AA}$ in both cases).

\subsection{HS Puppis = Nova Pup 1963}

\begin{figure}
\includegraphics[width=\columnwidth]{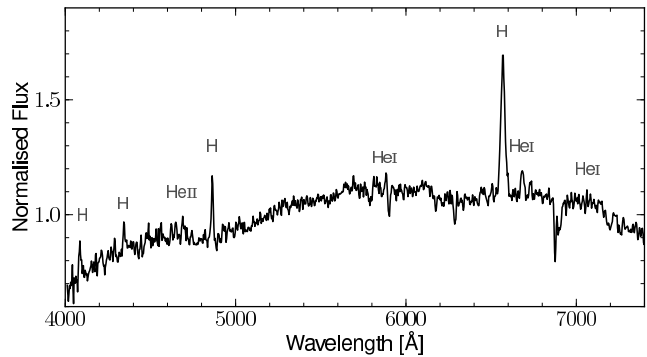}
\caption[]{Normalized low-resolution spectrum of HS Pup.
Positions of typical emission lines are indicated by corresponding labels.}
\label{hspupspec_fig}
\end{figure}

\begin{figure}
\includegraphics[width=\columnwidth]{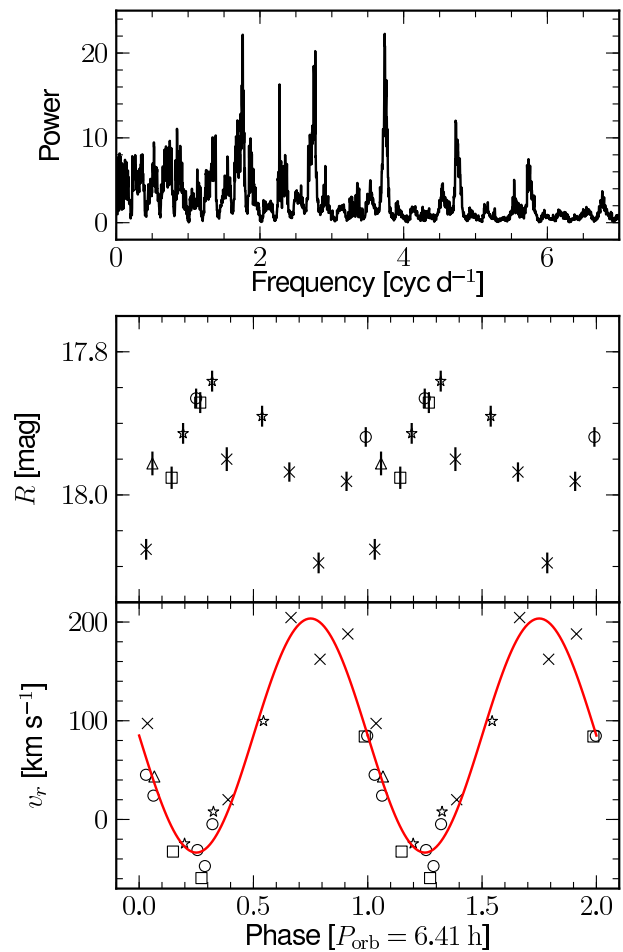}
\caption[]{Time-series data on HS Pup. Top: AOV periodogram of the radial
velocities of the H$\alpha$ peak component. 
Middle: $R$ light curve. Bottom: radial velocities of the H$\alpha$ peak
component. The sine represents the best fit. In the latter two plots, 
different symbols indicate different nights.}
\label{hspupdata_fig}
\end{figure}

Discovered by \citet{strohmeier64-6} on plates of the Bamberg Southern 
Station, it reached maximum on 1963 December 23 ($m_\mathrm{pg}$ = 8.0 mag). 
In spite of its southern position, its light curve could be monitored from the 
Sonneberg observatory until 1964 February \citep{huth+hoffmeister64-1}. 
\citet{duerbeck87-1} classifies it as a moderately fast nova with a decay
time $t_3$ = 65 d. \citet{gill+obrien98-1} detect a shell around HS Pup with 
a diameter of $<$2.5 arcsec. An optical spectrum can be found in 
\citet{zwitter+munari95-1}. As a single feature it presents a weak but rather
broad and potentially double-peaked H$\alpha$ emission line. Infrared 
magnitudes can be found in \citet{hoardetal02-1}. A time-series photometric
study was conducted by \citet{woudt+warner10-1}. They find strong flickering 
activity at time-scales of 15 min and amplitudes up to 0.15 mag. From a very 
weak periodic signal they suggest a possible orbital modulation with a period 
of 3.244 h and an amplitude $\Delta V = 0.0161~\mathrm{mag}$. Like many 
old novae, HS Pup undergoes low-amplitude long-term variations: 
\citet{szkody94-2} observed it in 1989 April at $V$ = 18.06 mag and 
\citet{woudt+warner10-1} found it at $V$ = 17.8 mag in 2000 December and 2001
February, as well as at $V$ = 18.0 mag in 2008 January. The only larger
deviation is reported by \citet{zwitter+munari95-1} who measured $V$ = 19.1 mag
from a spectrum taken in 1994 October.

The low-resolution spectrum shown in Fig.~\ref{hspupspec_fig} represents an
average of 12 spectra that were taken in 2009 May, with the system being at 
$R \sim$17.6 mag (Table \ref{obslog_tab}). In spite of measuring different 
$V$ magnitudes, \citet{szkody94-2} and \citet{zwitter+munari95-1} report
very similar colours of $V\!-\!R$ = 0.38 and 0.40 mag, respectively. This
yields $V \sim$18.0 for our spectrum. The much improved S/N with respect to 
the previously available data allows us to detect the full series of Balmer
emission lines, as well as the presence of He{\sc i}. Notable is the
weakness or even absence of the Bowen/He{\sc ii} $\lambda$4686 complex that
usually is one of the signatures of old novae. This suggests that the
mass-transfer rate in HS Pup is lower than in other post-novae. Comparison
with standard stars from several libraries 
\citep*{jacobyetal84-1,pickles85-2,silva+cornell92-1} shows that the red part 
of the continuum agrees well with that of an early to mid K star,
indicating a comparatively long orbital period.

Analysis of the radial velocities measured in spectra taken in 2011 
February/March indeed suggests several possible periods longward of 6 h: 
$P_1 = 6.41~\mathrm{h}$, $P_2 = 8.72~\mathrm{h}$ and $P_3 = 13.75~\mathrm{h}$
(Fig.~\ref{hspupdata_fig}, top). Folding the data on these periods yields
a significantly better fit for $P_1$ than for the other two. An additional
point in favour of $P_1$ is that it is approximately twice the photometric 
period detected by \citet{woudt+warner10-1}. We thus conclude that the orbital
period of HS Pup is $P_\mathrm{orb} = 0.2671(38)~\mathrm{d}
= 6.41(09)~\mathrm{h}$. Defining the point of the red-to-blue crossing
of the radial velocity curve as the zero-point of the orbital phase yields
the ephemeris to
\begin{equation}
T_0 ({\rm HJD}) = 245\,5624.7071(17) + 0.2671(38)~E~.
\end{equation}
The parameters corresponding to the best sine fit of the data are given
in Table \ref{rvpar_tab}. 
Using this sine fit within the framework of the time-series analysis shows
that the other peaks in the periodogram are aliases corresponding to the 
window function.

$R$ magnitudes measured from the acquisition frames indicate that the
system was about 0.3$-$0.4 mag fainter than for the 2009 spectrum. This is
also reflected in the emission lines. For the average 2011 spectrum we
measure the equivalent width of H$\alpha$ to 
$W_\mathrm{H_\alpha} = 25~\mathrm{\AA}$,
while in 2009 the line was slightly weaker ($W_\mathrm{H\alpha} = 
17~\mathrm{\AA}$).
Folding the $R$ magnitudes on the spectroscopic period yields a light curve
that is roughly consistent with an orbital modulation in the form of a 
sinusoidal variation with an amplitude of $\sim$0.25 mag. The apparent
presence of flickering, however, makes it necessary to average a much larger
number of measurements to analyse the shape of the light curve in detail.

\subsection{V909 Sagitarii = Nova Sgr 1941}

\begin{figure}
\includegraphics[width=\columnwidth]{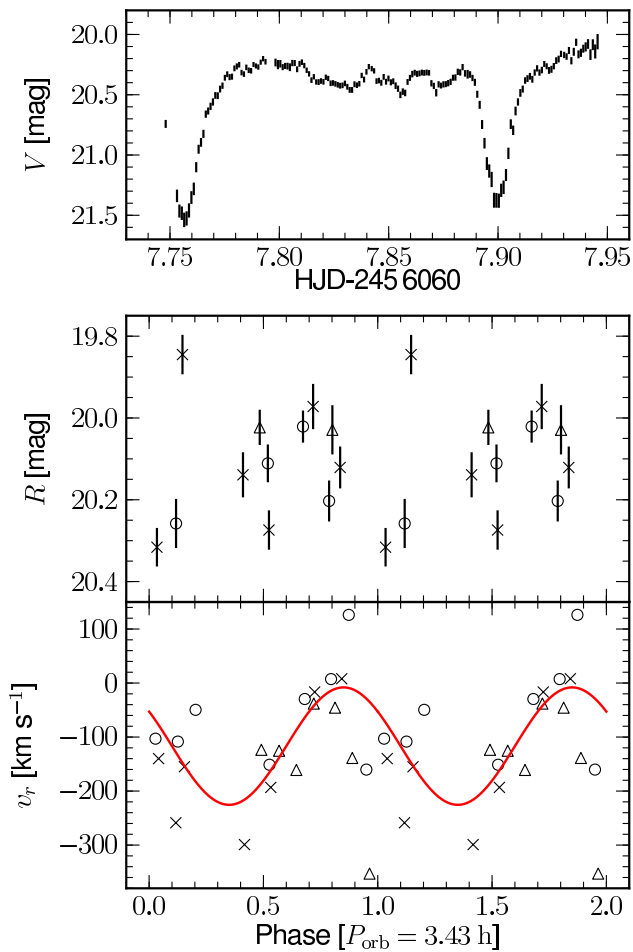}
\caption[]{Time-series data on V909 Sgr. Top: $V$ light curve. 
Middle: $R$ light curve. Bottom: radial velocities of the H$\alpha$ peak
component. The sine represents the best fit. In the latter two plots different
symbols indicate different nights.}
\label{v909sgrdata_fig}
\end{figure}

A brief history of this very fast nova \citep[$t_3$ = 7 d;][]{duerbeck87-1}
is included in \citetalias{tappertetal12-1}. The spectrum
presented there shows strong He{\sc ii} contribution indicating a magnetic
system, as already suspected by \citet{diaz+bruch97-1}. The latter authors
furthermore report the system to be eclipsing with an approximate period
of 3.36 h, but stated the need for confirmation.

Due to the faintness of the target the radial velocities of the H$\alpha$
emission line proved very difficult to measure, in spite of the
considerable strength of the line (Fig.~\ref{allspec_fig}; $W_\lambda$ =
41 {\AA}). We have therefore additionally taken a $V$-band light curve and
were fortunate to cover two eclipses with our data (Fig.~\ref{v909sgrdata_fig},
top). We measure the times of the minima of the two eclipses to 
HJD 245\,6067.75640(72) and HJD 245\,6067.89926(55). This yields the orbital 
period to $P_\mathrm{orb} = 0.14286(17)~\mathrm{d} = 3.4286(41)~\mathrm{h}$,
which corrects the preliminary value from \citet{diaz+bruch97-1} by about
four minutes. The formal ephemeris is thus
\begin{equation}
T_0 ({\rm HJD}) = 245\,6067.89926(55) + 0.14286(17)~E~.
\end{equation}
Note that the error estimation here does not take into account potential
influences of the timings by long-term changes. The two here recorded eclipses,
e.g., differ by $\sim$0.15 mag in depth. A longer time series would
thus be desirable to improve the statistics. 

Both the $R$ magnitudes from the acquisition frames and the H$\alpha$ radial
velocities show considerable noise when folded on above ephemeris. A
sine fit to the radial-velocity data yields the parameters summarized in
Table \ref{rvpar_tab}. The zero-point of the radial-velocity curve coincides
with a photometric phase of 0.10(02), indicating that the source of the
emission is approximately on the opposite side from the centre-of-mass as
seen from the secondary star.

Finally we briefly remark on the presence of a previously unidentified 
eclipsing variable star in the field of V909 Sgr. Its coordinates are 
RA$_\mathrm{2000.0}$ = 18:25:52.34, Dec.$_\mathrm{2000.0}$ = $-$35:03:05.4 
with an uncertainty of 0.25 arcsec. The average magnitude is $V$ = 18.7
mag and the light curve is reminiscent of a W UMa star. In our 4.74 h light 
curve we cover one eclipse and one maximum, the two having a brightness 
difference of $\Delta V$ = 1.04 mag. We determine the time of mid-eclipse to 
HJD 245\,6067.799(03). The time difference between the two features is 
2.45(10) h, so that the orbital period should be close to 9.8 h if the light 
curve is symmetric. This is a rather typical value for contact binaries
\citep[e.g.,][]{rucinski92-1}.

\subsection{V373 Scuti = Nova Sct 1975}

\begin{figure}
\includegraphics[width=\columnwidth]{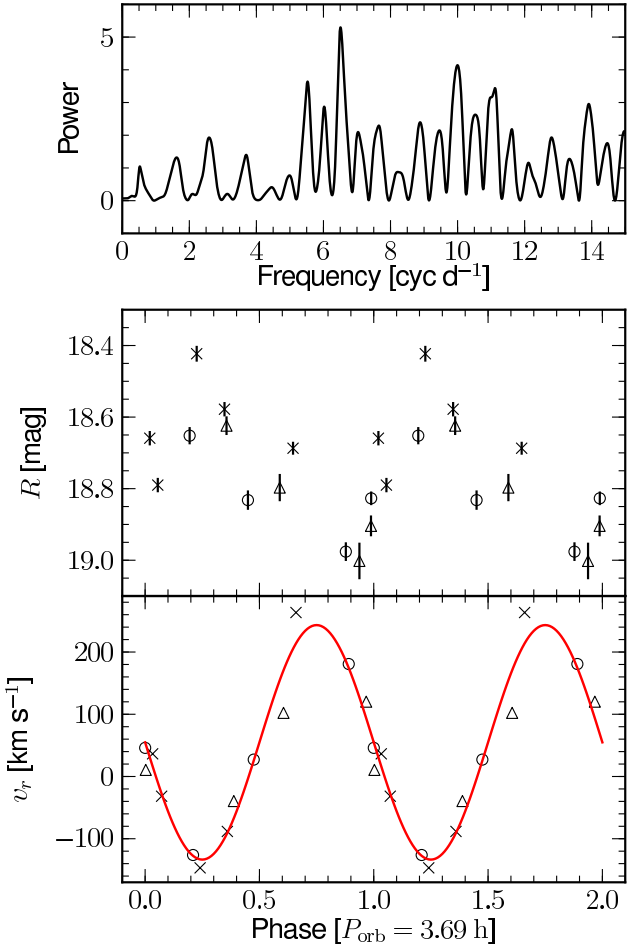}
\caption[]{Time-series data on V373 Sct. Top: Scargle periodogram of the
H$\alpha$ radial velocities. Middle: $R$ light curve.
Bottom: radial velocities of the H$\alpha$ line. The sine represents the best
fit to the data. In the latter two plots, different symbols indicate different
nights.}
\label{v373sctdata_fig}
\end{figure}

This moderately slow nova was discovered by \citet{wild75-4} at the Zimmerwald 
Observatory, Switzerland. Its eruption light-curve properties have been 
discussed by \citet*{stropeetal10-1} who assigned it subtype J with a decay 
time $t_3$ = 79 d and a maximum magnitude $V$ = 6.1. The subtype J (jitter) 
refers to post-maximum isolated flare-ups of the order of 1 mag. In V373 Sct,
these are observed only within the first $\sim$150 d after eruption. An 
optical spectrum of the post-nova taken by \citet{ringwaldetal96-3} shows 
strong Balmer emission and an equally strong Bowen/He{\sc ii} $\lambda$4686 
component. \citet{woudt+warner03-4} performed high-speed photometry of V373 
Sct, finding strong flickering activity at time-scales of 15 min and 
amplitudes up to 0.6 mag. They also found a coherent oscillation in one of 
their runs with a period of 258.6 s and suggest that V373 Sct might belong to 
the DQ Her type class of intermediate polars, with a rapid, non-synchronous 
rotation of the white dwarf. 

Our observations of V373 Sct cover three consecutive nights from 2011 June
29 to July 1 (Table \ref{obslog_tab}). The average spectrum 
(Fig.~\ref{allspec_fig}) presents a strong H$\alpha$ emission line with an 
equivalent width $W_\lambda$ = 53 {\AA}, which is slightly below the value of 
57 {\AA} recorded by \citet{ringwaldetal96-3}. Using their colour $V\!-\!R$ = 
0.3 mag, our $R$ magnitudes translate to $V \sim$18.4 mag. Thus, the system
was slightly brighter than in the \citet{ringwaldetal96-3} data ($V$ = 18.7
mag), which fits well with the weaker H$\alpha$ emission. Additionally,
the spectrum shows the clear presence of He{\sc i} $\lambda$6678 and
also a hint of very weak He{\sc i} $\lambda$7065.

Analysis of the radial velocities of the H$\alpha$ emission line yields
several possible alias peaks (Fig.~\ref{v373sctdata_fig}, top), but only the
highest one at $f = 6.51~\mathrm{cycle~d^{-1}}$ gives an acceptable 
radial-velocity curve. We thus identify this peak with the orbital period 
$P_\mathrm{orb} = 0.1536(28)~\mathrm{d} = 3.69(07)~\mathrm{h}$. The ephemeris
corresponding to the time of red-to-blue crossing results to
\begin{equation}
T_0 ({\rm HJD}) = 245\,5744.8376(10) + 0.1536(28)~E~.
\end{equation}

Fitting the radial velocities with a sine function yields an unusually large 
semi-amplitude $K_\mathrm{H\alpha} = 188~\mathrm{km~s^{-1}}$ 
(Fig.~\ref{v373sctdata_fig}, bottom; Table \ref{rvpar_tab}).
Consequently, the emission source has to be situated comparatively far
away from the centre-of-mass, with a distance of
\begin{equation}
a_\mathrm{H\alpha} \sin i = K_\mathrm{H\alpha} P_\mathrm{orb} / 2 \mathrm{\pi} 
\sim 0.57~\mathrm{R_\odot}
\end{equation} 
representing a lower limit. At an orbital period below 4 h thus either the 
system has a comparatively high mass ratio, or the main contributor to
the H$\alpha$ emission is (close to) the secondary star. In any case the
inclination of the system should be comparatively high. This is reflected
in the $R$ light curve that shows a variation with an amplitude of about
0.6 mag (Fig.~\ref{v373sctdata_fig}, middle). At first glance the light curve 
appears to have an unusual sawtooth shape with a short and steep rise and a 
slow decline. However, the incompleteness of the phase coverage leaves room 
for the possibility of an ellipsoidal shape with two minima of different 
depth at 0.5 and 0.9 orbits and two maxima of different height at phases 
0.2 and 0.7. Additionally, with the small number of data points the potential 
presence of flickering 
\citep[that had been observed in the $V$ filter by][]{woudt+warner03-4} can
distort the light curve significantly.

\subsection{CN Velorum = Nova Vel 1905}

\begin{figure}
\includegraphics[width=\columnwidth]{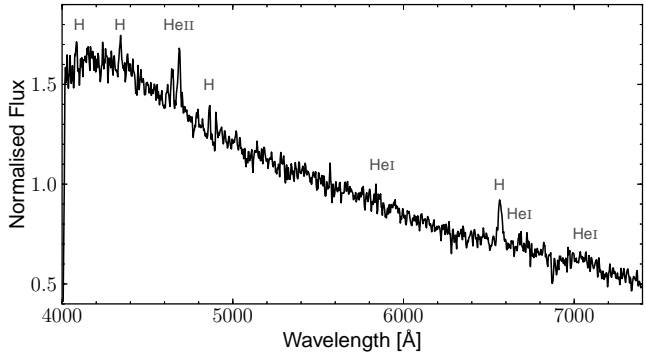}
\caption[]{Normalized low-resolution spectrum of CN Vel.
Positions of typical emission lines are indicated by corresponding labels.}
\label{cnvelspec_fig}
\end{figure}

\begin{figure}
\includegraphics[width=\columnwidth]{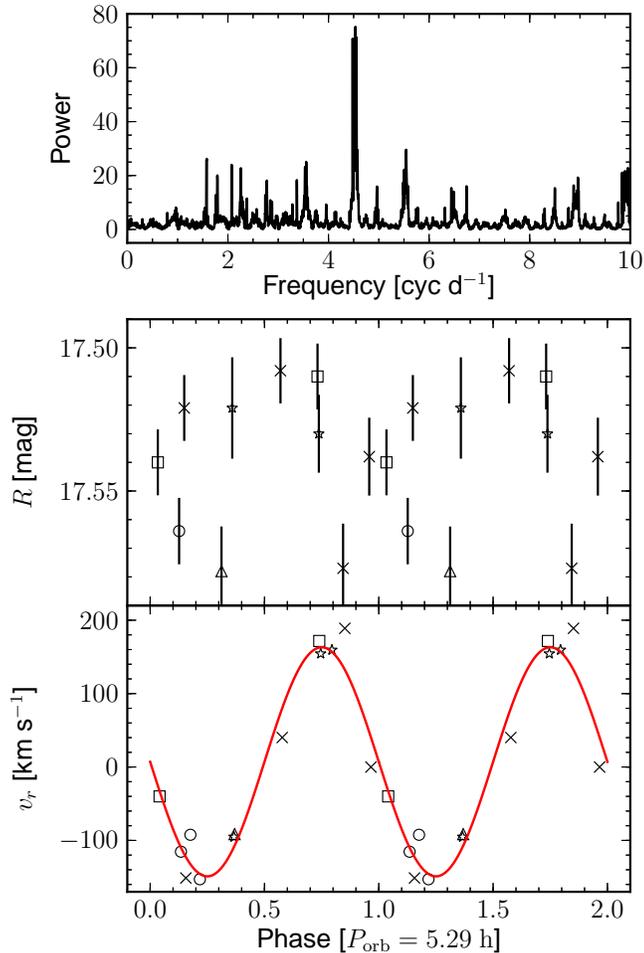}
\caption[]{Time-series data on CN Vel. Top: AOV periodogram of the
H$\alpha$ radial velocities. Middle: $R$ light curve.
Bottom: radial velocities of the H$\alpha$ line. The sine represents the best
fit to the data. In the latter two plots different symbols indicate different
nights.}
\label{cnveldata_fig}
\end{figure}

This very slow nova was discovered by Leavitt on Harvard plates 
\citep{leavitt+pickering06-4}. Peaking at $m_\mathrm{pg}$ = 10.2 mag, the
decay time was $t_3 > 800~\mathrm{d}$ \citep{duerbeck87-1}. A spectrum
of the post-nova is included in \citet{zwitter+munari96-1}. It shows a
noisy blue continuum with a weak H$\alpha$ emission line.

Our low-resolution spectrum in Fig.~\ref{cnvelspec_fig} presents a bit more 
detail. Apart from the Balmer series and a few weak He{\sc i} emission
lines the Bowen/He{\sc ii} $\lambda\lambda$4650/4686 component appears
particularly strong. Together with the steep blue continuum this indicates
that this post-nova is still maintaining a high mass-transfer rate.

Analysis of the radial velocities gives an unambiguous signal at 
$f = 4.55~\mathrm{cycle~d^{-1}}$ (Fig.~\ref{cnveldata_fig}, top) which we 
identify with the orbital period $P_\mathrm{orb} = 0.2202(30)~\mathrm{d} 
= 5.285(72)~\mathrm{h}$. As with V373 Sct, the amplitude of the 
radial-velocity curve is comparatively high (Fig.~\ref{cnveldata_fig}, bottom;
Table \ref{rvpar_tab}), which suggests that a significant amount of the
H$\alpha$ emission originates close to the secondary star. 
Defining the ephemeris by the means of the
red-to-blue crossing of the radial-velocity curve yields
\begin{equation}
T_0 ({\rm HJD}) = 245\,5624.8533(17) + 0.2202(30)~E~.
\end{equation}

In contrast to all the other systems in our sample, the photometric $R$-band
data of CN Vel do not appear to be modulated with the spectroscopic period
(Fig.~\ref{cnveldata_fig}, middle). However, we have seen that slow 
low-amplitude photometric variations in 
post-novae are a common phenomenon. In principle it could be possible that
the system was in a slightly lower state on the first nights of our 
observations, 2011 February 19 and 21. Excluding these two data points 
(the circle and the triangle) the remaining data appear consistent with
a sinusoidal signal of an amplitude of $\sim$0.06 mag. Note that this is
based on a mere assumption and needs confirmation by further observations.
The uncertainties of our photometric measurements are certainly large enough
to allow for any kind of variability or even none at all.
\citet{zwitter+munari96-1} measured $V = 17.78$ mag and $V\!-\!R$ = 0.16 mag 
for their observations. Applying their colour to our data yields $V \sim$17.7 
mag and thus a similar brightness.

CN Vel has a close visual neighbour, and while the finding chart in the
\citet{downesetal05-1} catalogue is unambiguous, the listed coordinates refer
to the combined light. We have measured the position of the nova on an
$R$-band image as described for AR Cir in Section \ref{arcir_sec} to
\begin{equation}
\alpha_{2000.0} = 11\!:\!02\!:\!38.66~~, 
\delta_{2000.0} = -54\!:\!23\!:\!09.5
\end{equation}
with a precision of $\sigma = 0.18$ arcseconds.

\section{Discussion and Conclusion}
\label{disc_sec}

\begin{figure}
\includegraphics[width=\columnwidth]{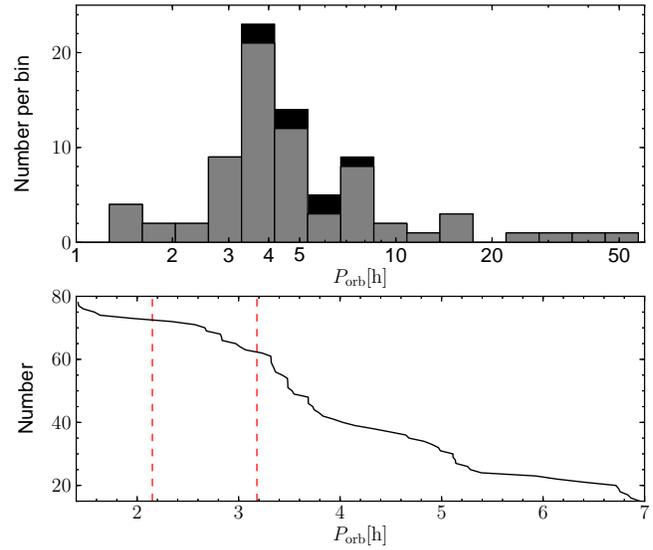}
\caption[]{Top: observed period distribution of classical novae on a
logarithmic scale. The data are from the \citet{ritter+kolb03-1} catalogue 
(grey) and the present paper (black). Bottom: the cumulative distribution
for periods 1.4--7.0 h. The dashed lines mark the limits of the period gap
from \citet{knigge06-2}.}
\label{rkp3hiscumu_fig}
\end{figure}

\begin{table}
\caption[]{Properties of the novae.}
\label{prop_tab}
\begin{tabular}{llllll}
\hline\noalign{\smallskip}
Object   & Year & $t_3$ & $\Delta m$ & $P_\mathrm{orb}$ 
& $W_\mathrm{H\alpha}$ \\
 & & (d) & (mag) & (h) & (\AA) \\
\hline\noalign{\smallskip}
V909 Sgr & 1941 & 7      & 13.5 & 3.43 & 41 \\
V373 Sct & 1975 & 79     & 12.6 & 3.69 & 53 \\
AR Cir   & 1906 & 415    & 8.2  & 5.14 & 19 \\     
CN Vel   & 1905 & $>$800 & 7.5  & 5.29 & 5  \\
V365 Car & 1948 & 530    & 8.0  & 5.35 & 5  \\
HS Pup   & 1963 & 65     & 10.0 & 6.41 & 25 \\
V972 Oph & 1957 & 176    & 8.5  & 6.75 & 2  \\
\hline\noalign{\smallskip}
\end{tabular}
\end{table}

\begin{figure}
\includegraphics[width=\columnwidth]{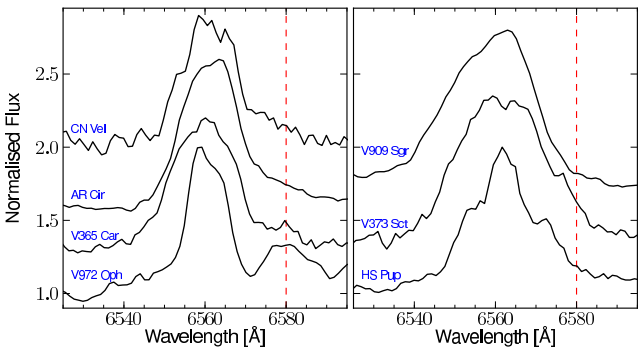}
\caption[]{H$\alpha$ line profiles of the average spectra that have been
normalized with respect to the maximum intensity of the line. The spectra
have been displaced vertically for display purpose. The dashed line marks the 
position of C{\sc ii}.}
\label{Haprof_fig}
\end{figure}

In the top plot of Fig.~\ref{rkp3hiscumu_fig} we present the updated period 
distribution on a logarithmic scale\footnote{The bin size is given by
$\Delta P = 10^{(n+1)/10}-10^{n/10}$ h, where $n = 0,1,2,...$} using the data 
from \citet[update 7.20, 2013 June]{ritter+kolb03-1}, limiting the
set to Galactic classical novae with unambiguous orbital periods, but
including the probable prehistoric novae Z Cam and AT Cnc
\citep[][respectively]{sharaetal07-1,sharaetal12-4}, to yield a total number of
71 orbital periods. The addition of the results of the present work thus
represents an increase of about 10 per cent, which emphasizes the undersampling
of the observed distribution. Following the measurement of the edges of the
period gap to 2.15 and 3.18 h by \citet{knigge06-2}, we can dissect the
period distribution into selected bins. We find that there are 6 systems 
(corresponding to 8 per cent) below the period gap and 10 (13 per cent) in the 
gap, with the large majority of systems occupying the range above the period
gap, but only 14 novae (18 per cent) with $P_\mathrm{orb} > 7~\mathrm{h}$. The 
fraction of novae in the 3--4 h range is 31 per cent and thus less than
the $\sim$50 per cent predicted by \citet{townsley+bildsten05-1}, and 
only two of our new additions fall into this regime. However, the cumulative 
distribution (Fig.~\ref{rkp3hiscumu_fig}, bottom) still shows that the slope
from $\sim$3.8--3.2 h is about a factor of 4 steeper than for other period
ranges, proving that this bin is overpopulated. The vast majority of 
CVs in this period range have been identified as SW Sex stars 
\citep{rodriguez-giletal07-2}, systems whose characteristics indicate that
they have very high $\dot{M}$ \citep*{rodriguez-giletal07-1}. This is thus
in agreement with the general idea of the importance of this parameter for
nova eruptions. It is furthermore confirmed that the slope within the
period gap is only slightly lower than e.g.~the range from 5 to 4 h, 
i.e.~there is a significant number of novae in the period gap. 
\citet{townsley+bildsten05-1} argue that these should contain a magnetic
white dwarf, in which case the interaction of the magnetic field with
the secondary star could prevent magnetic braking to operate 
\citep*{lietal94-1}. Evidence, albeit not always conclusive, for the presence
of a magnetic white dwarf has been found for V4633 Sgr 
\citep{lipkin+leibowitz08-1}, V351 Pup \citep{woudt+warner01-1}, 
V630 Sgr \citep{schmidtobreicketal05-2}, V2214 Oph \citep{baptistaetal93-1}, 
V597 Pup \citep{warner+woudt09-3}, V Per \citep*{woodetal92-2} and 
DD Cir \citep{woudt+warner03-4}, i.e.~for seven of the 10 novae in the gap. 
This exceeds the estimated fraction of $\sim$22 per cent of magnetic CVs 
\citep{araujo-betancoretal05-2} by far, and thus provides strong support for 
above hypothesis. 

In Table \ref{prop_tab}, we present selected properties of our novae. The year
of the eruption gives us the time $\Delta t$ that has passed from the eruption 
to our current observations, and thus the `age' of the post-nova. Next 
is the time $t_3$ in which the nova has declined by 3 mag from maximum 
brightness. The values were mostly taken from \citet{duerbeck87-1} and in one 
case (V373 Sct) from \citet{stropeetal10-1}. Column 3 gives the eruption 
amplitude $\Delta m$ which has been calculated as the difference between the 
reported maximum brightness and the average $V$ magnitudes of the post-nova. 
For the calculation of the latter we have included our own measurements as well
as data from the literature. Because the maximum magnitudes are not
corrected for the colour difference between $V$ and the (mostly) photographic 
plates the involved uncertainties can well be within the 0.5 mag range.
In spite of these uncertainties, our sample seems to present well the general 
behaviour of post-novae in the $\Delta m$ versus $\log t_3$ diagram, when 
considering all 11 novae contained in \citetalias{tappertetal12-1} (table 5) 
and in Table \ref{prop_tab} here. Within the expected scatter, they are in 
perfect agreement with the linear relation given in equation (1) of 
\citet{vogt90-1}, based on a much larger sample of post-novae. Column 4 gives 
the orbital period $P_\mathrm{orb}$ and column 5 gives the equivalent width 
$W_\mathrm{H\alpha}$ of the H$\alpha$ emission line, which shows a rough 
inverse dependency on $\dot{M}$ \citep{patterson84-1}. Note, however,
that such correlation is only valid if the majority of the H$\alpha$ emission
originates in the accretion disc and is thus indicative of its brightness.
In our sample we find two systems, where this is not the case: V909 Sgr,
which is a probable magnetic system, and HS Pup, whose line profile shows
a much stronger contribution of an additional component than in the
other systems (Fig.~\ref{Haprof_fig}). 

In Section \ref{v972oph_sec}, we have reported on the probable
detection of the C{\sc ii} $\lambda\lambda$6578,6583 doublet in emission.
Carbon emission is a common feature in the ultraviolet spectra of CVs
\citep[e.g.,][]{ladous91-3}, and as part of the Bowen blend at $\lambda$4640
{\AA} in high $\dot{M}$ CVs \citep[e.g.,][]{williams83-1}. Additionally,
a few CVs show enhanced carbon emission, mainly in the blue part of their
spectra \citep[e.g.,][]{drewetal03-1,schmidtobreicketal03-5,bianchinietal12-1}.
The carbon content in CVs and pre-CVs is a potential indicator of the 
evolutionary status of the binary and especially the secondary star, which
has motivated several corresponding studies 
\citep*{gaensickeetal03-1,harrisonetal04-1,harrisonetal05-1,%
harrisonetal05-2,tappertetal07-2,howelletal10-1,hamiltonetal11-1}. 
However, to our knowledge this particular C{\sc ii} emission has never been
observed before in a CV with the exception of V445 Pup, where it originates
in the expelled nova shell \citep{iijima+nakanashi08-1}, while in V972 Oph
it is clearly situated in the binary itself. A favourable factor in its
detection is certainly the comparatively narrow and weak H$\alpha$ emission
line. This raises the question if the presence of (perhaps slightly weaker)
C{\sc ii} in other CVs is not simply masked by the typically broad and
strong H$\alpha$ line. In Fig.~\ref{Haprof_fig}, we compare the line profiles
of the seven novae. To account for the different line strengths we have
normalized the spectra with respect to the maximum intensity of the line.
On the left-hand side of the plot we have collected the `carbon suspects',
i.e.~the novae that show a certain extra flux near 6580 {\AA}. 
Apart from V972 Oph these are V365 Car, AR Cir and CN Vel. The line profiles 
of the remaining three novae in the right-hand plot of the figure show more 
symmetric wings, although the line in V373 Sct is certainly sufficiently broad 
to hide potentially present weak C{\sc ii}. The inclusion of AR Cir in this
list demonstrates that a red extended H$\alpha$ emission wing can also
indicate the presence of an outflow rather than C{\sc ii} emission. 
Considering that evidence for such outflow has been detected for several
CVs in ultraviolet lines \citep[e.g.,][]{froning05-2} but also in H$\alpha$ and
He{\sc i} \citep{kafka+honeycutt04-2}, this represents perhaps an even more 
likely explanation. Additionally, the outflow phenomenon in CVs appears to
be most commonly found in systems with high $\dot{M}$ and most post-novae 
certainly fall into that category. However, assuming that such outflows are
bipolar \citep{drew87-1} we would then expect to find a blueshifted counterpart
either in absorption (for an optically thick outflow) or in emission
(for an optically thin outflow as has been recently detected in VY Scl;
Schmidtobreick et al., MNRAS, submitted). Apart from AR Cir, in none of
the other targets any such component can be detected in our data, but on the
other hand a (weak) absorption component could be easily masked by the 
principal H$\alpha$ emission line.

The data presented here were obtained with the intention to determine the
orbital periods of the seven novae, and they served this purpose well.
A closer inspection reveals interesting phenomena in a number of systems,
but the limited quality of the data mostly allows only a glimpse at them
and often raises more question than answers. Several of the novae here merit
further investigation, and we hope that the present work motivates more
detailed studies.

\section*{Acknowledgements}
We thank Elena Mason and Antonio Bianchini for discussion on the spectrum
of V972 Oph and Roberto Saito for comments regarding the VVV infrared data
on several novae.

This research was supported by FONDECYT Regular grant 1120338 (CT and NV).
AE acknowledges support by the Spanish Plan Nacional de Astrononom\'{\i}a y 
Astrof\'{\i}sica under grant AYA2011-29517-C03-01. 

We gratefully acknowledge ample use of the SIMBAD data base, 
operated at CDS, Strasbourg, France, and of NASA's Astrophysics Data System 
Bibliographic Services. {\sc iraf} is distributed by the National Optical 
Astronomy Observatories.

\label{lastpage}

\end{document}